\documentclass[12pt]{article}

\usepackage{amsmath, amsfonts, amssymb}
\usepackage{graphicx}

\def\be{\begin{equation}}
\def\ee{\end{equation}}
\def\ba{\begin{eqnarray}}
\def\ea{\end{eqnarray}}

\def\bdm{\begin{displaymath}}
\def\edm{\end{displaymath}}
\def\la{~\mbox{\raisebox{-.6ex}{$\stackrel{<}{\sim}$}}~}

\def\bq{\begin{quote}}
\def\eq{\end{quote}}

 at 10truept

\newcommand{\beq}{\begin{equation}}
\newcommand{\eeq}{\end{equation}}
\newcommand{\bea}{\begin{eqnarray}}
\newcommand{\eea}{\end{eqnarray}}
\newcommand{\beqa}{\begin{eqnarray}}
\newcommand{\eeqa}{\end{eqnarray}}

\newcommand{\al}{\alpha}
\newcommand{\bt}{\beta}
\newcommand{\gam}{\gamma}

\newcommand{\del}{\delta}
\newcommand{\Del}{\Delta}
\newcommand{\eps}{\epsilon}

\newcommand{\sig}{\sigma}

\newcommand{\om}{\omega}

\newcommand{\de}{\partial}
\newcommand{\rmd}{\mathrm{d}}

\renewcommand{\[}{\left[}
\renewcommand{\]}{\right]}
\renewcommand{\(}{\left(}
\renewcommand{\)}{\right)}

\newcommand{\n}{\hat{n}}

\renewcommand{\L}{\mathcal{L}}

\newcommand{\cm}{\textrm{cm}}
\newcommand{\eV}{\textrm{eV}}
\newcommand{\GeV}{\textrm{GeV}}
\newcommand{\TeV}{\textrm{TeV}}

\newcommand{\GHz}{\textrm{GHz}}
\newcommand{\nG}{\textrm{nG}}
\newcommand{\Mpc}{\textrm{Mpc}}
\newcommand{\kpc}{\textrm{kpc}}
\newcommand{\erg}{\textrm{erg}}

\newcommand{\pp}{\parallel}

\def\ltap{\ \raise.3ex\hbox{$<$\kern-.75em\lower1ex\hbox{$\sim$}}\ }
\def\gtap{\ \raise.3ex\hbox{$>$\kern-.75em\lower1ex\hbox{$\sim$}}\ }
\def\gl{\ \raise.5ex\hbox{$>$}\kern-.8em\lower.5ex\hbox{$<$}\ }
\def\roughly#1{\raise.3ex\hbox{$#1$\kern-.75em\lower1ex\hbox{$\sim$}}}

\begin{document}

\thispagestyle{empty}
\begin{flushright}
January 2015
\end{flushright}
\vspace*{1.5cm}
\begin{center}
{\Large \bf Anisotropies in Non-Thermal Distortions of }\\
\vspace*{.3cm} 
{\Large \bf Cosmic Light from Photon-Axion Conversion}\\

\vspace*{1.25cm} {\large Guido D'Amico$^{a, }$\footnote{\tt
damico.guido@gmail.com} and Nemanja Kaloper$^{b, }$\footnote{\tt
kaloper@physics.ucdavis.edu}}\\
\vspace{.5cm} {\em $^a$Center for Cosmology and Particle Physics, New York University, New York, NY 10003}\\
\vspace{.5cm} {\em $^b$Department of Physics, University of
California, Davis, CA 95616}\\

\vspace{1.5cm} ABSTRACT
\end{center}
Ultralight axions which couple sufficiently strongly to photons can leave imprints on the sky at diverse frequencies by mixing with cosmic light in the presence of background magnetic fields.
We explore such direction dependent grey-body distortions of the CMB spectrum, enhanced by resonant conditions in the IGM plasma. We also find that if such axions are produced in the early universe and represent a subdominant dark radiation component today, they could convert into X-rays in supervoids, and brighten them at X-ray frequencies.

\vfill \setcounter{page}{0} \setcounter{footnote}{0}
\newpage

Axions are hypotetical pseudo-scalar particles motivated by -- to date -- the only known dynamical solution of the strong CP problem. In the Peccei-Quinn mechanism \cite{Peccei:1977hh}, an anomalous symmetry is spontaneously broken at a high scale. The phase of the complex field which picks a {\it vev} is the Goldstone mode, which gets a potential from nonperturbative QCD corrections, so that at the minimum the residual CP-violating phase is tiny, $\theta_{CP} \le 10^{-9}$ \cite{Weinberg:1977ma, Wilczek:1977pj}. UV completions of the Standard Model, provided by string theory constructions, give rise to many more such phases, which at low energies behave as ultralight fields. One linear combination couples to QCD, and gets a mass from the QCD nonperturbative effects. Others may remain much lighter, and therefore active at cosmological scales. If they couple to long range $U(1)$ fields, in particular to electromagnetism, they could yield interesting astrophysical effects, providing extra dimming of supernovae, 
distorting CMB and affecting quasar spectra and cosmic rays -- see for example \cite{Anselm:1986gz}-\cite{Csaki:2014ona}. Even if they are decoupled from gauge theories, but only couple gravitationally, owing to their lightness they can still leave astrophysical signatures, in particular due to the phenomenon of superradiance as explained in \cite{Arvanitaki:2009fg, Arvanitaki:2010sy}. 

Clearly, searching for axions is easier when there are direct axion-matter couplings. These couplings are given by irrelevant operators in order 
to maintain shift symmetry, and thus are suppressed by some high mass scale. However in the presence of nonvanishing background fields these couplings can be enhanced by the background and yield possibly observable effects. For example, if one such ultralight axion couples to photons, by the standard  
dimension-5 operator $\L_{a-\gam} = \frac{ \phi}{4M} \tilde{F}_{\mu \nu} F^{\mu \nu}$, in the presence of background magnetic fields this will induce a bilinear term $\L_{a-\gam \, \textrm{eff}} = \frac{B_{\rm background}}{M} \phi E$ that yields 
photon-axion oscillations. Depending on the coherence length of the background field, distance travelled and the environmental conditions, an incoming flux of photons can be partially converted to axions and evade detection. What remains, however, will -- in general -- have a different distribution over the range of frequencies than it had at the source, since the mixing rate is frequency dependent. Hence if a source of photons were well understood, seeking for the spectral distortions can either indirectly reveal the presence of a photon-axion mixing, or constrain it. Conversely, it is also possible that in the presence of relic axions, the conversion can in fact enhance a flux of photons, making certain sections of the sky where the mixing is strong brighter than the others. From a fixed observer's perspective, this would make the spectral distortions line-of-sight dependent, and so anisotropic. 

These phenomena were applied to the search for ultralight axions, with masses $\le 10^{-15} \eV$. The reason is that in cosmological environments, the presence of the plasma, with plasma frequency induced effective photon mass, $m_\gamma \ge 10^{-15} \eV$, very efficiently screens the mixing. Similarly a large axion mass would screen it even more efficiently. However these screening effects may cancel, enhancing  the mixing to close to maximal even for low frequencies. This could happen if the plasma frequency is sufficiently close to the axion mass in large regions of space, where the plasma is sufficiently smooth. Candidates for such regions, which should be quiet, are very large voids. In light of the fact that such supervoids have been recently discovered~\cite{Szapudi:2014zha,szapgarcia}, we will consider what happens with the photons and axions inside them, if the magnetic fields are large and the plasma sufficiently slowly varying. Also, we will consider the mixing effects in the clusters, where the plasma frequency is higher than in the voids, but the magnetic fields can also be much larger. As long as the magnetic coherence length is comparable to the length scale controlling the variation of the plasma density, the conditions for resonant mixing may be met. For this reason we will consider axions with 
masses in the range  $10^{-12} - 10^{-16} \; \eV$, with the mass  being a scanning parameter, such that 
the resonant conditions may be met somewhere in space. This can occur in the framework of the axiverse, with a plenitude of light axions \cite{Arvanitaki:2009fg}. Of course, it is still necessary that the axion which can be in resonance is the one coupling to the usual electromagnetism.

Assuming this occurs, we will show how the mixing can produce a \emph{direction-dependent} grey-body distortion of the CMB spectrum. While our results are to be viewed as estimates, they are an interesting indicator of the potential of cosmological observations for discovering axions in this as yet largely unexplored regime of parameters, where the observations are rapidly improving. So, for example, a dedicated analysis on the Planck data can provide a map of the regions through which photons have disappeared into axions\footnote{
We are aware that this analysis may be difficult because of calibration uncertainties.
However, future experiments dedicated to measuring the CMB spectrum will be sensitive to the photon disappearance.}.

Second, we will consider the possibility that there is a residual high frequency axion background, as suggested in \cite{Conlon:2013isa}. The premise of this scenario is that the universe contains axion `dark radiation', generated by a late decay of heavy string moduli, with a spectrum peaked today at the soft X-ray frequencies. These axions would convert into X-ray photons in background magnetic fields inside clusters.  This yields an excess X-ray luminosity from galaxy clusters \cite{Conlon:2013txa, Kraljic:2014yta, Powell:2014mda}, claimed to match some recent observations \cite{Angus:2013sua}.
While the uncertainties in the predicted X-ray luminosity at present limit the direct discovery potential of such searches, we point out that a possible tie-breaker could be provided by voids. Namely, inside the magnetic fields in voids, which are weaker but possibly more coherent, 
the relic axions could convert relatively efficiently into photons, 
making the void brighter in the X-ray part of the spectrum.
Such a signal might be present in the Rosat X-ray and XMM-NEWTON data, and a dedicated analysis therefore seems warranted.

The axion-photon mixing dynamics is described by the 
action
\be
S = - \int \rmd^4 x \sqrt{-g} \[ \frac{1}{2} \de_\mu \phi \de^\mu \phi + \frac{1}{2} m_a^2 \phi^2
+ \frac{1}{4} F_{\mu \nu} F^{\mu \nu} + \frac{1}{4 M} \phi \tilde{F}^{\mu \nu} F_{\mu \nu} \] \, ,
\ee
where $F_{\mu \nu}$ is the standard electromagnetic tensor, $\tilde{F}^{\mu \nu} = \frac{1}{2 \sqrt{-g}} \eps^{\mu \nu \rho \sig} F_{\rho \sig}$, $m_a$ is the axion mass and $M$ is a mass parameter controlling the coupling strength. 
We will be picking $M$ near the current limits, $M \simeq 10^{11} ~ \GeV$ \cite{cast}, since our main goal is to illustrate the sensitivity of the current observations to the effects induced by ultralight axions. 
So we assume that the couplings and environmental magnetic fields are as strong as possible while not being excluded by observations at present.
The future data may constrain these couplings more tightly. 
We ignore the space-time curvature effects because we will work with the axion masses $m_a$ much greater than the relevant curvature scales. Furthermore, the flavor mixing can only take place in relatively low plasma density environments with significant background magnetic fields, 
at least of the order of nano-Gauss, but with coherent lengths much shorter than the curvature scales. Such conditions might only occur at very low redshifts $\le 2$, and so the cosmic evolution is largely negligible. This also implies, that any effects from photon-axion mixing will be suppressed for very old light, since at such early times the plasma frequencies were higher and magnetic fields likely negligibly small according to current ideas on magnetogenesis.

In the presence of a background magnetic field $B$, the equations yield a simple bilinear system which describes photon-axion oscillations~\cite{Anselm:1986gz, Yoshimura:1987ma, Raffelt:1987im}.
The conversion probability of a beam of photons which travels a distance $z$ in the presence of a magnetic field $B$ admits a perturbative solution 
\be
P_c = \left| \int_0^z \rmd s \, e^{-i \Del_a s} e^{i \int_0^s \rmd r \Del_p(r)} \mu_\pp(s) \right|^2 \, ,
\ee
where $\Del_a = m_a^2/(2 \om)$, $\Del_p = \om_p^2/{2 \om}$, $\mu_\pp = B_{\pp}/M$, and $\om_p^2 = 4 \pi \al n_e/m_e$ is the plasma frequency, with $n_e$ the number density of free electrons. In the integral above we allow for spatial variations of the magnetic field $B$ and the electron number density $n_e$, but treat it adiabatically as in
\cite{Raffelt:1987im}. 

To gain physical insight into the dynamics, we can specialize to the case of constant coherent magnetic field filling a spatial domain of size $L_{dom}$ and containing a constant electron number density $n_e$.
Solving the coupled system of field equations governing photon-axion propagation \cite{Csaki:2001yk}, 
we find the standard transition probability of a 2-level quantum system:
\be
P_t = \frac{\mu^2 L_{dom}^2}{4} \( \frac{\sin(\eta L_{dom})}{\eta L_{dom}} \)^2 \, ,
\label{eq:Pconv}
\ee
where $\eta$ is the inverse of the oscillation length, $\eta = \frac{\sqrt{(\om_p^2 - m_a^2)^2 + 4 \mu^2 \om^2}}{4\om} $.
Depending on the parameters and the frequency of the modes, several different regimes can set in:
\begin{enumerate}
\item {\it maximal mixing}, with $\omega \gg \frac{|\omega_p^2-m_a^2|}{2\mu}$ everywhere in the space where photons propagate, and the probability of transitions $P_t \simeq \sin^2(\mu L_{dom}/2)$ is independent of frequency regardless of the size of magnetic domains;
\item  {\it non-maximal mixing, where $\eta L_{dom} \ll 1$}, which means that for the fixed frequency of interest the magnetic domain is very thin: in this case, $P_t \simeq \mu^2 L_{dom}^2/4$ is constant and frequency-independent;
\item {\it non-maximal mixing, with $\eta L_{dom} \gg 1$}, implying that for a fixed frequency the magnetic domain is very thick: in this case, the oscillations can be averaged and we end up with
$P_t \simeq \mu^2/(8 \eta^2) \simeq 2 \mu^2 \om^2/(\om_p^2 - m_a^2)^2$, which has a $\om^2$ frequency dependence, and a small amplitude; the crossover frequency between the latter two regimes is 
$\omega_{crossover} \simeq \frac{(\omega_p^2 - m_a^2)L_{*}}{2}$, where $L_* = \min(L_{dom}, M/B)$ \cite{Csaki:2014ona};
\item \emph{resonant regime}, where in some regions $|\om_p^2 - m_a^2|$ happens to be very small compared to 
$2 \mu \om$ locally, due to the fact that the plasma density changes to be close to the axion mass.
In this case, inside this region the conditions for maximal mixing are fulfilled and frequency dependence is very weak locally; the photon survival probability inside only such regions is $P_t \simeq \sin^2(\mu L_{dom}/2)$, while outside is given by the formulas governing case 2) or 3) \cite{Yoshimura:1987ma, marshsilk}.
\end{enumerate}
So axions will mix with photons of any frequency at some level in background magnetic fields, with the 
transition rate depending on the frequency.
This has possible important cosmological implications for the Cosmic Microwave Background photons (CMB) and for the high energy cosmic ray photons. We will explore them below, focusing in fact on the case
4) for both CMB photons and soft X-rays in clusters, and case 1) for high energy X-rays.
Note, of course, that in a realistic situation the actually observed flux of photons will be the result of propagation through many different domains along the line of sight.
This means that the total rate of photon-axion conversions will be determined by a combination of different dynamical regimes, possibly involving -- in general -- a combination of some, or even all, of the cases listed above, due to the random variation of environmental parameters, including strength and direction of background magnetic fields, plasma density, residual axion abundance and so on.
For example, the photon-axion conversion in the optical range, which can affect supernovae, is a consequence of integrating over many randomly oriented domains where some of the conditions listed above apply, as described in \cite{Csaki:2001yk}.
For CMB, and in the environments of interest to us in what follows, however, the transitions in individual domains are suppressed unless the special conditions allowing faster transition rates occur. Hence we will ignore the contributions from many domain in the rest of this article, focusing exclusively on the cases where one domain along the line of path contributes dominantly to photon-axion conversion.

\begin{figure}
\centering
\includegraphics[scale = 0.5]{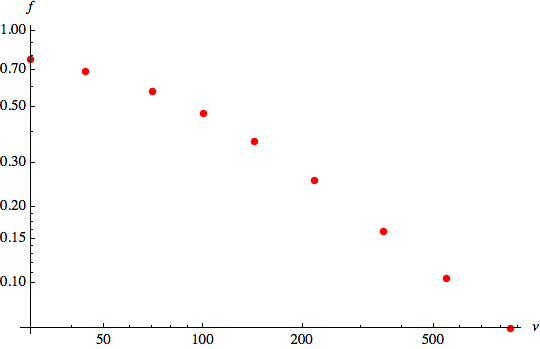}
\caption{Plot of the function $(1-e^{-x})/x$, as a function of the frequency in $\GHz$. The points correspond to the central values of the Planck frequency bands.}
\label{fig:spectrum}
\end{figure}

CMB has a thermal spectrum to a very good accuracy because of tight coupling with baryons before recombination. This form of the distribution is preserved by the adiabatic expansion of the Universe.
The temperature varies very little with direction, because inhomogeneities were small at recombination.
Perturbations in the local density of photons can be interpreted as a local variation in the black-body temperature:
\be
I_{BB}(\om, \n) = \frac{2 \hbar \om^3}{c^2} \[ \exp\( \frac{\hbar \om}{k_B T(\n)} \) - 1 \]^{-1} \, ,
\label{bbspect}
\ee
where $I_{BB}(\om, \n)$ is the radiation spectrum as a function of energy $E = \hbar \omega$ and angular direction $\n$. If there is any local variation of the temperature, there will be a local excess or lack of photon energy density, given by the leading order variation of (\ref{bbspect})
\be
\del I(\om, \n) = \left. \frac{\de I}{\de \ln T} \right|_{T_0} \frac{\del T(\n)}{T_0}
= I_{BB} \frac{x}{1 - e^{-x}} \frac{\del T(\n)}{T_0} \, ,
\ee
where $x \equiv \hbar \om/(k_B T_0)$ and $T_0 = 2.726 \, {\rm K}$ is the direction-averaged temperature measured by COBE-FIRAS.
Since in the presence of background magnetic fields and axions CMB photons can disappear, at a rate which depends in general on their frequency, and varies with the line of sight, local spectral distortions can be induced.
Indeed, the spectrum of photons surviving the propagation from the last scattering surface to the detector is
\be
I_{\rm c}(\om, \n) =  I_{BB}(\om) (1 - P_c(\n, \om)) \, ,
\ee
which obviously deviates from a pure black body spectrum. 
Converting this spectral distortion into thermodynamic temperature units, we get
\be
\frac{\del T_{\rm conv}}{T} = - P_c(\n, \om) \frac{1 - e^{-x}}{x} \, .
\ee
This shows that the thermodynamic temperature acquires a frequency dependence even if the conversion probability is independent of frequency -- as in the case 4) listed above, which we will focus on since if it is realized for the CMB photons and realistic weak intergalactic magnetic fields, it yields the dominant distortions of the spectral distribution.
The cases 1) and 3) can never yield significant effects for CMB photons with realistic parameters  taken into account. The case 2) is more interesting, but also generically small without resonance conditions (however see \cite{Csaki:2014ona} for possible thermal distortions in extragalactic space which are potentially discernible by comparison of several different frequency channel data sets).The resultant frequency dependence of the distorted spectrum due to this effect is shown in Fig.~\ref{fig:spectrum}.

Let us now demonstrate that the scale of the effect is within the reach of observations, as claimed above.
The weakest assumption in a part of our argument is that the magnetic fields should be sufficiently strong.
Since they are not known directly we will assume that they are close to saturating the current limits.
A justification of this assumption is that while the size of the effects we are discussing might not be viewed as a firm prediction, it {\it is} a qualitative prediction of a type of an imprint in the CMB. If observed, such effects could be independent checks not only of the axion's presence but also of the extragalactic magnetic fields, which are otherwise very hard to observe directly. Alternatively, the effect could also be enhanced even if the magnetic fields are weaker, if the resonance conditions (case 4) above are met in some sufficiently large and smooth region of space with coherent magnetic fields. This at first sounds like fine tuning of the Lagrangian parameters -- i.e., the axion mass $m_a$ -- to the environmental parameters -- i.e., the plasma frequency $\omega_p$ \eqref{eq:etacluster}. One may worry if this can ever be achieved. A viable way to relax this criticism is the paradigm of the axiverse~\cite{Arvanitaki:2009fg} which posits that there exist many light axions, with masses that are small and that scan the full range of cosmologically interesting scales, down to the current value of the Hubble parameter $H_0$. In such a framework, the resonant conditions are realized along some directions for at least some of the axions. The signatures which we discuss can arise if that particular axion is the one which couples to the usual electromagnetism.

With this in mind, we first consider the photon-axion conversion in clusters. There occur the coherent magnetic fields $B \sim \mu {\rm G}$, with coherence length $\simeq \rm kpc$.
This implies that the $\mu$ parameter is
\be
\mu = \frac{B_\pp}{M} = 6.9 \times 10^{-28} \( \frac{B_{\parallel}}{\mu {\rm G}} \) \( \frac{10^{11} \GeV}{M}\)\eV \, .
\ee
The plasma frequency can be estimated as
\be
\om_p =\sqrt{\frac{4 \pi \al}{m_e} n_e } = 5.5 \times 10^{-13} x_e^{1/2} \( \frac{\Del}{10^3} \)^{1/2} \eV  \, ,
\ee
where $x_e$ is the ionization fraction of electrons ($x_e \simeq 1$ at low redshifts), and $\Del \equiv \rho_m/\bar{\rho}$ is the local overdensity parameter.
For a cluster, we expect $\Del \simeq 10^3 - 10^4$.
The angular frequency for a typical microwave photon of frequency $\nu_{100} \equiv \nu /100 \, \GHz$ is
$\om = 4.1 \times 10^{-4} \, \nu_{100} \, \eV $.
Using these values, and defining $B_X = B_\pp/X$, $M_{11} = M/10^{11} \GeV$, the (square of the) inverse oscillation length will be (with axion mass $m_a \rightarrow m_a/\eV$ in what follows)
\begin{multline}
\label{eq:etacluster}
\eta^2 = \frac{(\om_p^2 - m_a^2)^2}{16 \om^2} + \frac{\mu^2}{4}
= \[ \( \frac{3 \times 10^{-28} \Del - m_a^2}{1.7 \times 10^{-3} \nu_{100}}\)^2 + 1.2 \times 10^{-55} \(\frac{B_{\mu G}}{M_{11}} \)^2 \] \eV^2\\
= \[ (\Del - 3.3 \times 10^{27} m_a^2)^2 \frac{3.1 \times 10^{-50}}{\nu_{100}^2} + 1.2 \times 10^{-55} \(\frac{B_{\mu G}}{M_{11}} \)^2  \] 2.4 \times 10^{52} \, \kpc^{-2} \, .
\end{multline}
Since the product $\eta L_{dom}$ is generically large, the photon-axion mixing is very suppressed 
away from the resonance. At the resonance, however, the transition probability is
\be
P_{t, res} = \sin^2 \( \frac{\mu}{2} L_{dom} \) \simeq 2.9 \times 10^{-3}
\(\frac{B_{\mu G}}{M_{11}} \)^2 \( \frac{L_{dom}}{\rm kpc} \)^2 \, ,
\ee
which in fact is even larger than the observed CMB anisotropies! Clearly, this sounds like a dangerously large effect. But 
with slightly weaker and less coherent magnetic fields, with weaker couplings, and less uniform plasma the conversion rates are suppressed and
the signatures are weaker. In fact one would expect this anyway, because the variation of the plasma density spoils the resonant conditions. 
Also, the conversion factor to thermodynamic temperature is always less than unity (Fig. \ref{fig:spectrum}).

We note that there is another interesting possibility where the photon-axion transitions can occur through voids, which are actually the largest fraction of the volume of the Universe.
This can happen if the magnetic fields are large enough; assuming the values $B \simeq \nG$, $L_{dom} \simeq \Mpc$ and an underdensity $\Del \simeq 10^{-1}$, we get
\begin{multline}
\eta^2 = \frac{(\om_p^2 - m_a^2)^2}{16 \om^2} + \frac{\mu^2}{4}
= \[ \( \frac{3 \times 10^{-28} \Del - m_a^2}{1.7 \times 10^{-3} \nu_{100}}\)^2 + 1.2 \times 10^{-61} \(\frac{B_{\nG}}{M_{11}} \)^2  \] \eV^2\\
= \[ (\Del - 3.3 \times 10^{27} m_a^2)^2 \frac{3.1 \times 10^{-50}}{\nu_{100}^2} + 1.2 \times 10^{-61} \(\frac{B_{\nG}}{M_{11}} \)^2  \] 2.4 \times 10^{58} \, \Mpc^{-2} \, .
\end{multline}
As in the case of clusters, $\eta L_{dom} \gg 1$, and again the effect should happen in resonance in order to maximize it. Since the plasma density is about 
$\om_{pl} = 1.2 \times 10^{-14} \sqrt{\frac{n_e}{10^{-7} \cm^{-3}}} \eV$, the resonant regime can occur for light axions,
with masses $m_a \la 10^{-14} \eV$. While the required resonance regime is narrower in the parameter space, it can nevertheless occur in random voids.
The size of the transition probability can actually be of the same order as for clusters:
\be
P_{t, res} = \sin^2 \(\frac{\mu}{2} L_{dom} \) \simeq 2.9 \times 10^{-3}
\(\frac{B_{\nG}}{M_{11}} \)^2  \( \frac{L_{dom}}{\rm Mpc} \)^2 \, .
\ee
So as a consequence, the thermal distortions could be imprinted as a signature of voids bearing large magnetic fields. Again, the scale of the transition probability is larger than the observed CMB anisotropies, but that is a consequence of our assumptions about the strength of photon-axion couplings, magnetic field strengths and coherence, and plasma homogeneity scales. In reality, these can be all weakened yielding smaller but possibly observable effects.

Finally, we stress that to be observable for realistic parameters, the spectral distortions from the photon-axion conversions should be distinguishable from the distortions due to the foregrounds.
The foreground effects are typically expressed in the units of antenna (or brightness) temperature, defined in the Rayleigh-Jeans approximation, which yields  
\be
T_A(\om, \nu) = \frac{c^2}{2 k_B \om^2} I(\om, \nu) \, .
\ee
In the antenna temperature units, the distortion due to axion conversion translates into
\be
\frac{\del T_A}{T} = - P_c(\n, \om) \frac{x}{e^x-1} \, .
\ee
A measurement of such spectral distortions may be feasible with the actual  
Planck data~\cite{Ade:2013ktc}, and would provide a map of photon ``disappearance'' along the line of sight.
The possible degeneracy with foregrounds, in particular synchrotron at low frequencies, may be resolved by noting that the spectral index of the axion-induced corrections $\bt \simeq - 0.3$ around 30 $\GHz$ is quite different from the synchrotron distortions, and so the foreground effects may be subtracted away at least in principle.

Similar phenomena also arise for very high energy photons. These photons -- i.e. cosmic X-rays and $\gamma$-rays  -- could be less sensitive to electron plasma since their frequency is so high. In extragalactic space, they are in the regime of maximal mixing, i. e. their dynamics falls in the category of the case 1) we listed above. In clusters, where the plasma density is higher and magnetic fields are stronger, the transition between the maximal and non maximal mixing can occur at higher frequency. This is further exacerbated if the photon-axion coupling is weaker as considered by 
\cite{Conlon:2013isa}. However, we note that if the axion mass is in the resonant regime relative to the plasma, as described by the case 4) which we listed above, the transitions can be enhanced yielding a stronger signal. As we noted above, the possibility of photon-axion resonance in plasma can be naturally realized in the context of the axiverse where the axion mass parameter scans a range of scales, and so it could hit the right value to be in resonance in some regions of space. This makes the X-ray signatures in both extragalactic space and in clusters possibly sensitive to the presence of such axions.

In more detail, the enhanced X-ray luminosity 
can occur if there is a relic  cosmic axion background (CAB), of very high energy axions.
Specifically, \cite{Conlon:2013isa} proposed a scenario where such relic background axions arise as a subdominant dark radiation from a late decay of heavy string moduli dark matter. The estimated density 
of CAB axions in \cite{Conlon:2013isa}  is of the order
\be
\rho_{CAB} = 1.6 \times 10^{60} \frac{\erg}{\Mpc^3} \frac{\Del N_{eff}}{0.57} \, ,
\ee
where $\Del N_{eff}$ is the fractional contribution of the axions to the effective number of relativistic species in addition to photons and Standard Model neutrinos.
Such radiation amounts to only $10 \%$ of the CMB energy density, however with the peak around $\rm keV$ energies.
As argued in \cite{Conlon:2013txa}, in the presence of a background magnetic field these axions will convert into soft X-ray photons. The expression for conversion probability is of course eq.~\eqref{eq:Pconv}. The parameters which the authors of \cite{Conlon:2013txa} employ in their analysis correspond to the dynamics of case 2), with marginally non-maximal mixing and frequency sensitivity. 

However, as we noted above, if the resonant regime occurs for the conversion of CMB photons, it can also occur for the conversion into X-ray photons. Obviously, this will not be realized inside every cluster, or over the whole volume of a cluster where the resonance may take place. Nevertheless, since the environmental conditions 
vary from cluster to cluster, and the axion mass can scan the regime of scales, the resonant conditions can be satisfied {\it somewhere}.
The conversion rate per axion per second will be $r \simeq P_{conversion}\, c/L_{dom}$. Using our previous estimates of the photon-axion conversion probabilities, we can estimate the total X-ray luminosity density of a cluster where resonant conversion sets,
\be
\mathcal{L} = \rho_{CAB} \, r \simeq  4.5 \times 10^{46}  \(\frac{B_{\mu G}}{M_{11}} \)^2  \( \frac{L_{dom}}{\rm kpc} \) \(\frac{\Del N_{eff}}{0.57}\) \frac{\erg}{\rm s \, \Mpc^3}  \, .
\label{lumicluster}
\ee
The overall scale of the excess luminosity in (\ref{lumicluster}) by itself is \underbar{too large}. However, as before with the imprints on CMB, it diminishes with weaker photon-axion couplings and less coherent plasma and magnetic fields. Specifically, since the total X-ray luminosity of a typical cluster due to conventional electromagnetic processes is $\L \sim 10^{44} \, \erg \, {\rm s}^{-1}$, taking the weaker photon-axion couplings
$M_{11} \simeq {\cal O}(1000)$, and without assuming resonance, the authors  of~\cite{Conlon:2013txa} conclude that conversion of the CAB axions into photons can yield an observable X-ray luminosity excess if the relic axion abundance is sufficiently high, and even point to certain reported astrophysical anomalies.
However, because of astrophysical uncertainties, it is harder to firmly establish that the axions are the cause for the excess luminosity noted in those cases~\cite{Kraljic:2014yta}. Nevertheless, the possibility remains, and as we see above, the resonance conditions -- if realized -- can yield effects with even weaker couplings.

Here we also point out another intriguing possibility for the detection of high energy CAB axions:  instead of looking at the clusters, look at the voids. The point is that although in the voids 
the magnetic field is weaker, the axion density is parametrically the same, since they are ultra-relativistic.
With nano-Gauss magnetic fields in large voids with the resonant conditions satisfied in regions of size $L_{dom} \sim \Mpc$, 
we can estimate the luminosity by scaling down the formula (\ref{lumicluster}), using the fact that the magnetic fields are a thousand times weaker while
the domain is a thousand times larger. This yields
\be
\mathcal{L}^{void} = \rho_{CAB} \,  r \simeq  4.5 \times 10^{43} \(\frac{B_{\nG}}{M_{11}} \)^2  \( \frac{L_{dom}}{\Mpc} \)  \(\frac{\Del N_{eff}}{0.57}\) \frac{\erg}{\rm s \, \Mpc^3}  \, ,
\ee
which is also very large. Again, as we stressed above, this is not a problem of principle, because the luminosity drops when the magnetic fields are weaker, the resonant domains are smaller than the whole $\simeq \Mpc$ sized underdensity, and the axion relic abundance is smaller.  Nevertheless our 
point that this is a possible signature of the photon-axion mixing in the voids remains. This could be a new portal into the dark sector of the universe,  and deserves more attention, both theoretical and observational.

Yet another interesting signature of photon-axion conversion, and evidence of axions,  is at yet higher energies, for $\gamma$-rays. These can be generated in high energy processes in astrophysical sources such as AGNs.
In the standard picture, the spectrum of $\gamma$-rays drops sharply at energies $\gtrsim 1 \TeV$ because of pair production in collisions with extra-galactic background light. Previously, the photon-axion mixing mechanism at such high frequency has been considered to allow ultra-high cosmic $\gamma$-rays from very distant astrophysical sources to surpass the GZK cutoff limits \cite{cktpgzk}.
However, as suggested in~\cite{DeAngelis:2008sk, Wouters:2012qd}, photon-axion mixing could harden the observed high-energy spectrum.
Indeed, as explained in~\cite{cktpgzk}, high energy $\gamma$-rays will partly convert into axions in the strong magnetic field close to the source, and travel undisturbed to our Galaxy, where they will convert back to photons and reach our detectors. While the astrophysical uncertainties obscure the precise identification of such events, it may be possible to get to them statistically, by systematically searching for anisotropies in the spectra of AGN's.
In particular, \cite{Wouters:2012qd} proposes exploring correlation between sources on small scales and an anti-correlation on large scales, which is a common characteristic of different magnetic field models.
While the tests carried out on current Fermi data are inconclusive at the moment, the future observations by the 
Cherenkov Telescope Array may have enough statistics to test part of the parameter space of photon-axion mixing and the presence of light axions which couple to photons.

To summarize, we have considered observable effects of photon-axion mixing in cosmological conditions.
We have argued that the photon-axion mixing can leave an imprint in the CMB, especially for the case where the axion mass -- as in the axiverse -- could happen to have the right value in order to be in resonance with plasma frequency in voids.
Future explorations of CMB may be sensitive to such imprints, as already argued in~\cite{Csaki:2014ona}, where the case was made that transitions out of resonance could yield distortions of the CMB. Here we pointed out that distortions of the overall temperature arise even when the dominant photon-axion conversions are frequency independent, just because the survival probability of a CMB photon is changing the normalization of the total black body radiance.
Such effects may be the strongest in the voids, and given that now very large voids appear to be an experimental fact 
\cite{Szapudi:2014zha,szapgarcia}, it seems warranted to look for such effects. It may be possible to determine such temperature shifts by the frequency-dependent analysis of the observed spectra. We also noted that the voids could be glowing in the X-ray spectrum if there is a relic cosmic axion background as well. Given the difficulties in directly searching for weakly coupled dark sectors such as axions, these cosmological windows may be a promising direction to explore seriously.

\vskip.8cm
	
{\bf Acknowledgments}: 
We thank Csaba Cs\'aki, Joseph Gelfand, Andrei Gruzinov, Raphael Flauger, Matt Kleban, Giulia Migliori, John Terning and Ingyin Zaw for useful discussions. N.K. is grateful to CCPP, New York University, for kind hospitality during the completion of this work. N.K. is supported in part by the DOE Grant DE-SC0009999.
G. D'A. is supported by a James Arthur Fellowship.

\end{document}